# Beyond Earthly Limits: Protection against Cosmic Radiation through Biological Response Pathways

Zahida Sultanova[1]*, Saleh Sultansoy[2]†

[1]School of Biological Sciences, University of East Anglia, Norwich, UK

[2]Tobb University of Economics and Technology, Ankara, Turkey

*Corresponding Author: zahida.sultanova@uea.ac.uk

†Member of ATLAS, LHeC & FCC collaborations at CERN

## Abstract

The upcoming phase of space exploration not only includes trips to Mars and beyond, but also holds great promise for human progress. However, the harm caused by cosmic radiation, consisting of Galactic Cosmic Rays and Solar Particle Events, is an important safety concern for astronauts and other living things that will accompany them. Research exploring the biological effects of cosmic radiation includes experiments conducted in space itself and in simulated space environments on Earth. Notably, NASA's Space Radiation Laboratory has taken significant steps forward in simulating cosmic radiation by using particle accelerators and is currently pioneering the progress in this field. Curiously, much of the research emphasis thus far has been on understanding how cosmic radiation impacts living organisms, instead of finding ways to help them resist the radiation. In this paper, we briefly talk about current research on the biological effects of cosmic radiation and propose possible protective measures through biological interventions. In our opinion, biological response pathways responsible for coping with stressors on Earth can provide effective solutions for protection against the stress caused by cosmic radiation. We also recommend establishing *the Dedicated International Accelerator Laboratory for Space Travel related radiation research (DIAL-ST)* to advance this field and evaluate protective biological pathways through particle accelerator experiments simulating cosmic radiation.

# 1. Introduction

Future space missions to destinations like the Moon, Mars and beyond are expected to cause significant health risks to astronauts and other organisms, mainly due to the damage caused by cosmic radiation (Chancellor et al., 2014). Unlike Earth, which shields its inhabitants with a protective magnetic field and atmosphere, space offers no such defence (Heinrich et al., 1999). As a result, astronauts and other space-traveling organisms, which evolved under Earth's protection, are vulnerable to the diverse and harmful components of cosmic radiation (Furukawa et al., 2020; Hagen, 1989). Because of the lack of natural resistance to cosmic radiation, advanced protective measures are necessary to protect the organisms that will travel through space.

The harmful cosmic radiation includes two primary sources: galactic cosmic rays and solar particle events and they are composed of protons (89%), alpha particles (9%), heavy ions (1%) and electrons (1%) (Piazzoli et al., 2022; Rauch et al., 2021). The percentage of heavy ions which have high energy is relatively low but their biological effects are expected to be more detrimental because their interaction with matter is proportional to their charge (Rauch et al., 2021). The energy spectrum of cosmic radiation exhibits a broad peak in the range of about 1 GeV (Piazzoli et al., 2022). Beyond this, the flux of cosmic rays decreases in proportion to an increasing power of their energy by $e^{-2.7}$ (Piazzoli et al., 2022). During space missions, organisms are routinely exposed to these radiation sources. Particularly, missions aboard the International Space Station (ISS) result in radiation doses much higher and more dangerous than those experienced on Earth, and further missions are predicted to involve even higher doses (Thirsk et al., 2009).

Researchers have begun investigating the effects of cosmic radiation on living organisms by using two primary approaches. The first approach is to send these organisms directly into space to study radiation effects in the space environment. By doing so, these organisms face not only natural cosmic radiation but also microgravity in locations like the International Space Station (but not on planetary surfaces such as Mars) (Kiefer & Pross, 1999). However, this method is challenging due to the cost and difficulty of sending living organisms into space. Moreover, it is hard to determine whether the observed effects are caused by cosmic radiation or microgravity (Kiefer & Pross, 1999; Yatagai et al., 2019). The second method is exposing organisms to cosmic radiation here on Earth. This is achieved by simulating cosmic radiation through particle accelerators that are able to produce electron and ion beams (Kuzmic et al., 2019; Norbury et al., 2016; Yi et al., 2013). Although both these approaches help us understand how cosmic radiation damages living things, studies exploring protection against cosmic radiation are relatively scarce. This scarcity is largely due to the novelty of this field and the limited research on biological interventions (such as drugs, diets and supplements) that can protect organisms from cosmic radiation.

In this article, we propose that one potential valuable defence against cosmic radiation lies in how organisms respond to natural stresses on Earth, such as elevated heat or starvation. In the next section, we discuss how cosmic radiation causes harm to organisms. Then, in section 3, we briefly talk about the biological interventions (such as drugs and dietary supplements) that have been studied for protection against cosmic radiation. In section 4, we explore which protection mechanisms organisms have evolved against natural stressors and how they can also be used for protection against cosmic radiation. In section 5, we look upon the particle accelerator facilities that are used for exploring cosmic radiation effects on biological organisms. Finally, we propose that multidisciplinary collaboration is necessary to study how biological stress response mechanisms can be studied to protect against cosmic radiation.

## 2. The Impact of Cosmic Radiation on Living Things

Cosmic radiation has well-documented harmful impacts on living organisms (Figure 1) (Furukawa et al., 2020; Kennedy & Wan, 2011). Among these, the most extensively investigated effect is severe DNA and chromosomal damage (Arena et al., 2014). Such damage can occur through both direct and indirect mechanisms. Direct damage happens when radiation directly harms genomic components (Montesinos et al., 2021). Indirectly, cosmic radiation can generate reactive oxygen species which are highly reactive chemicals that in high concentrations, can harm the genome (Brieger et al., 2012; Montesinos et al., 2021). Moreover, beyond genomic harm, cosmic radiation can also adversely affect the structure and function of other biomolecules, including proteins and lipids (Fang et al., 2002;

Radman, 2016; Shaler et al., 2020). The detrimental effects of cosmic radiation on DNA and other biomolecules have far-reaching impacts on diverse aspects of life, spanning lifespan, reproduction, stress resistance, social/cognitive abilities and cancer risk (Alaghband et al., 2023; Furukawa et al., 2020; Ikenaga et al., 1997; Kuzmic et al., 2019; Lenarczyk et al., 2023; Sakashita et al., 2010; Trani et al., 2010; Yi et al., 2013). Consequently, cosmic radiation constitutes a substantial concern for nearly all organisms destined for space exploration.

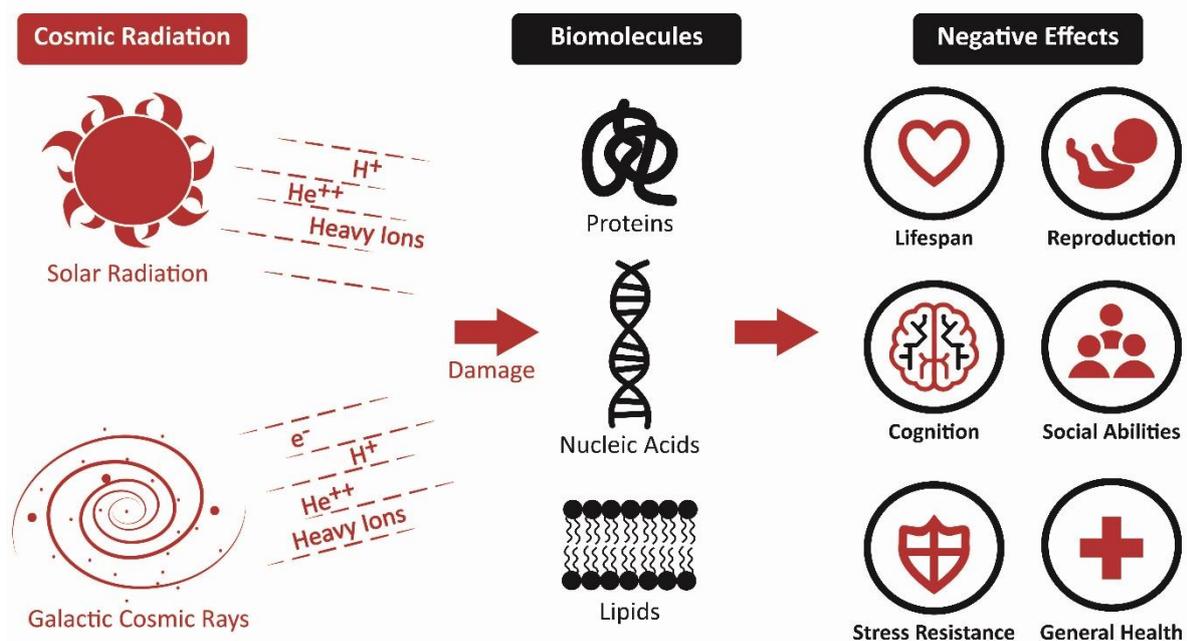

**Figure 1. Negative Effects of Cosmic Radiation.** Cosmic radiation consists of Solar Particle Events and Galactic Cosmic Rays. It contains protons ($H^+$), alpha particles ($He^{++}$), heavy ions and electrons ($e^-$) which can damage biomolecules that are essential for the survival of the living things. This damage can lead to many negative effects in lifespan, reproduction, cognition, social abilities, stress resistance and general health.

Model animals such as *Caenorhabditis elegans* (worms), *Drosophila melanogaster* (fruit flies) and mice as well as model plants such as *Arabidopsis thaliana* (thale cress) and *Brassica rapa* (field mustard) have been used for investigating the negative effects of cosmic radiation. For example, in worms, heavy ion irradiations resulted in lower reproduction by halting germ cell proliferation (Sakashita et al., 2010). Likewise, high energy protons reduced resistance to oxidative stress in worms (Yi et al., 2013). In fruit flies, cosmic radiation during space travel led to the accumulation of harmful mutations (Ikenaga et al., 1997). Moreover, fruit flies that were sent to space experienced adverse effects in reproduction and immunity, although it was not possible to separate the effects of cosmic radiation and microgravity in that experiment (Iyer et al., 2022). Additionally, mice exposed to cosmic radiation simulation at NASA's cosmic radiation laboratory had social and cognitive impairments (Alaghband et al., 2023; Kiffer et al., 2022), cardiovascular problems (Lenarczyk et al., 2023) and higher incidence of tumours (Trani et al., 2010). While plants tend to be more radiation-resistant compared to animals, they are not immune to the damage from cosmic radiation (Arena et al., 2014). *Arabidopsis* plants exposed to accelerated carbon ions, for instance, suffered from irreparable mutations (Hase et al., 2012). Likewise, subjecting seeds of both *A. thaliana* and *B. rapa* to simulated cosmic radiation had negative effects on their viability and root length (Zhang et al., 2022). These studies collectively emphasize the potential harm caused by cosmic radiation to animals and plants intended for future space missions.

Research about the effects of cosmic radiation on human health comes from studies on astronauts that are back from space travels (Rose, 2022). Although there are also studies exploring the effects of radiation therapy and nuclear radiation on humans, they are not ideal for understanding cosmic radiation effects because their dose and duration is different from the radiation experienced on space (Kennedy & Wan, 2011). Astronauts studies have shown that even relatively small doses of cosmic radiation can cause important health problems (Cucinotta et al., 2001; Delp et al., 2016). For example, cosmic radiation has been associated with early onset of cataracts (Cucinotta et al., 2001) and a higher risk of cardiovascular disease (Delp et al., 2016). It is also suggested that cosmic radiation might lead to increased cancer risk (Cucinotta, 2024; Guo et al., 2022). However, confirming a clear correlation between space travels and cancer risk is challenging due to the small sample size of astronauts (Elgart et al., 2018). It has been proposed that cosmic radiation can also harm human fertility by negatively affecting reproductive cells (Barbrow, 2020; Straume, 2015). While this problem has not been studied in astronauts, it is suggested that cosmic radiation may reduce a woman's ovarian reserve by up to 50% during a mission to Mars (Rose, 2022). Altogether, these human studies underline the importance of further research how to protect the health of astronauts on future space missions.

## 3. Biological Interventions for Safe Space Travel

Protecting against cosmic radiation involves various methods such as physical shields and biological defences (Furukawa et al., 2020). While physical shields have been traditionally used to provide protection in space (Naito et al., 2020), they are not effective enough to reduce exposure to cosmic radiation (Durante, 2014). Moreover, high energy particles can go through the shields and interact with the components of the shield. Thus can cause extra radiations that can be even more detrimental for living things (Durante & Cucinotta, 2008). Therefore, increasing the organism's natural biological defence can provide an additional layer of protection against the harmful components of cosmic radiation.

By far, the biological interventions that were tested are mainly drugs and dietary supplements that reduce the damage caused by radiation exposure (Guan et al., 2004, 2006; Xie et al., 2023). For example, CDDO, which is an antioxidant/anti-inflammatory drug, can reduce the cancerogenic effects of heavy ion beams in mice (Suman & Fornace, 2022). Similarly, antioxidants, including several vitamins, minerals and amino acids, can be used as dietary supplements to protect against reactive oxygen species and inflammation that result from cosmic radiation (Montesinos et al., 2021). In fact, an antioxidant combination that includes vitamin E and vitamin C protected mice from the oxidative stress caused by proton, heavy ion and gamma radiation (Guan et al., 2006). While these interventions show promise, they constitute just one piece of the puzzle. We can discover more ways to protect against cosmic radiation, by studying how animals handle other types of stresses on Earth. In other words, Earthly stress-response mechanisms can offer helpful ideas for safeguarding against space stress.

## 4. Biological Response Pathways against Cosmic Rays

Organisms do not experience primary cosmic rays on Earth because Earth's geomagnetic field and atmosphere are able to deflect and de-energize charged particles arriving from space. (Heinrich et al., 1999). Nevertheless, organisms on Earth have to deal with various other stressors such as food/energy deprivation or heatwaves (Burchfield, 1979; Hightower, 1991; Lindquist, 1986; White, 2008; Yiming,

2006). In response to these stressors on Earth, organisms activate specific biological response pathways such as nutrient/energy sensing and stress response pathways (Efeyan et al., 2015; Richter et al., 2010; Sultanova et al., 2023). They are generally considered as adaptive responses because they not only protect the organisms from external damage but also can save energy until the stressful situation is over (Efeyan et al., 2015; Richter et al., 2010; Sultanova et al., 2021). We propose that these adaptive responses can also provide protection against the damage caused by space radiation (Figure 2).

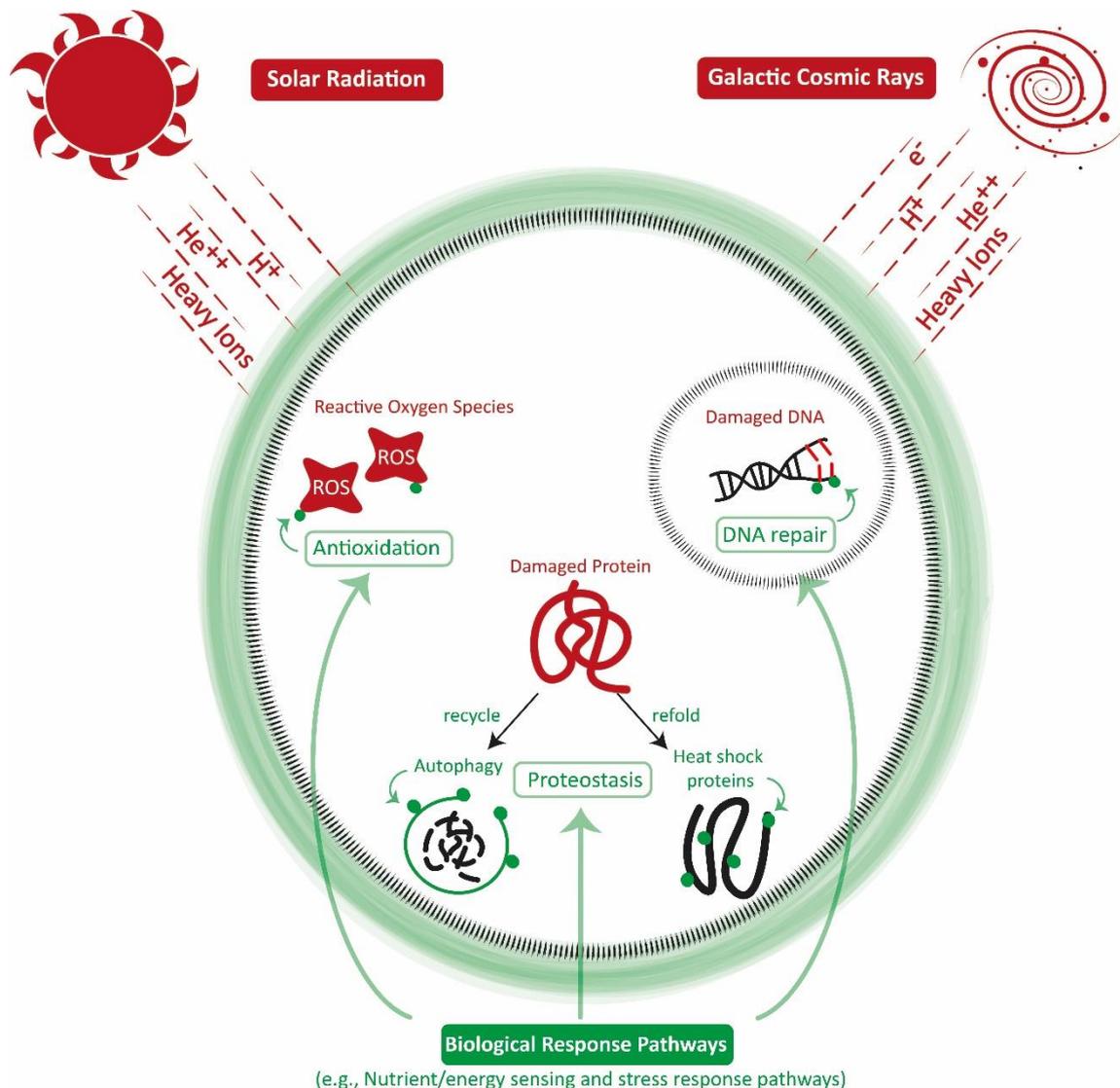

**Figure 2. Biological Stress Response Pathways Against Cosmic Radiation.** Cosmic radiation from the sun (Solar Particle Events) and beyond (Galactic Cosmic Rays) can harm living organisms by increasing reactive oxygen species (ROS) and damaging important biomolecules like DNA and proteins. However, biological pathways that evolved to protect against Earthly stressors can reduce this harm through several mechanisms. Antioxidation can removes ROS, DNA repair enzymes fix DNA damage and proteostasis recycles damaged protein fragments or refolds misfolded proteins. These protective mechanisms can be enhanced by modulating biological response pathways through interventions such as mutants, drugs or dietary manipulations. Damage caused by cosmic radiation is shown in red while the protective mechanisms are shown in green.

Nutrient sensing pathways, such as insulin/IGF-1 signaling and TOR pathways, detect the presence of food (Efeyan et al., 2015). Hence, when these pathways are inactive, an organism transitions to a conserving state until food becomes available again. Inactivating them stimulates DNA repair mechanisms, increases resistance to reactive oxygen species and strengthens protein homeostasis (Chávez et al., 2007; Conn & Qian, 2011; Hyun et al., 2008). One way to maintain protein homeostasis is autophagy, a process that recycles non-functional proteins and organelles (Y. Y. Chang et al., 2009; Lum et al., 2005). This process is conserved in eukaryotic cells from yeast to animals and plants and is necessary for clearing and reusing damaged cellular components (Bassham et al., 2006). Likewise, energy sensing pathways, such as AMPK and NAD/Sirtuin pathways, are activated when there is energy limitation (H. C. Chang & Guarente, 2014; Hardie et al., 2016). Thus, when they are active, an organism enters a protective stage until the return of the energy. This results with decreased protein synthesis, increased autophagy and better antioxidant response (Bolster et al., 2002; Meijer & Codogno, 2011). Finally, stress response pathways, like the heat shock response, are triggered when a stressor, such as heat, is present (Lindquist, 1986). Heat stress leads to misfolded and dysfunctional proteins, subsequently activating heat shock response pathways (Morimoto et al., 1997). These pathways are involved in correctly folding misfolded proteins and recycling the non-functional proteins with the help of heat shock proteins and autophagy (C. Hu et al., 2022; Morimoto et al., 1997). Remarkably, the heat shock proteins are even more conserved than autophagy, as they can be found in simple bacterial cells as well as complex eukaryotic cells like animals and plants (Sørensen et al., 2003). Considering the damaging effects of cosmic radiation on DNA, proteins, and lipids and its tendency to increase reactive oxygen species (ROS) production, all these biological pathways are expected to play a critical role in protection against cosmic radiation.

Significantly, these biological pathways can be modulated by molecular or pharmacological interventions in addition to the presence of the actual stressor (Dillin et al., 2002; Fornace & Suman, 2022; D. Hu et al., 2021; Li et al., 2014). By using mutant strains or specific drugs that target these pathways (Figure 2), we can explore how these pathways can be used for protection against cosmic rays. In fact, metformin, which is a drug known to activate the AMPK energy sensing pathway, suppresses heavy ion-induced tumorigenesis in mice (Fornace & Suman, 2022). However, this is just the initial phase of exploration. Exploring these biological pathways is valuable to understand how biological interventions can enhance protection against the deleterious effects of cosmic radiation. Notably, these interventions have already been studied in model organisms for treatments against diseases such as cancer and diabetics as well as for extending healthy lifespan (Cifarelli et al., 2015; Harrison et al., 2009; Martin-Montalvo et al., 2013). Some of the drugs targeting those pathways, such as metformin and rapamycin, are already FDA approved and are being used by humans, making their study relatively straightforward and cost-effective (Corcoran & Jacobs, 2018; Lamanna et al., 2011; Li et al., 2014).

## 5. Particle Accelerators as Cosmic Radiation Simulators

Three global centres are mainly used for investigating the effects of cosmic radiation on biological systems. The most well-known one is NASA's Space Radiation Laboratory that contains a Galactic Cosmic Ray simulator which provides 33 sequential ion beams (energies from 50 MeV/u to 1.5 GeV/u) (Huff et al., 2023; Norbury et al., 2016). This facility has been used to investigate the impact of space radiation on many different biological characteristics, including cancer progression, heart function, cognitive performance and behaviour (Huang et al., 2019; Kiffer et al., 2022; Nemec-Bakk et al., 2023; Raber et al., 2020). Second, the GSI Helmholtz Center for Heavy Ion Research can provide a wide range of particles from hydrogen to uranium (energies from 100 MeV/u to 2 GeV/u) (Durante et al., 2010). Recently, this centre was used to explore the effects of heavy ion exposure on brain cells (Roggan et al., 2023). Third, the Heavy Ion Medical Accelerator in Chiba (HIMAC) provides ion beams such as helium, carbon, argon, iron and xenon (energies from 100 MeV/u to 800 MeV/u) (*HIMAC*, 2006; Miller, 2019). Lately, the effect of proton and Fe ion beams on chromosomal damage on embryonic stem cells was investigated in this centre (Yoshida et al., 2022). While diverse particle beams can be sequentially used in these centres, exposing organisms to different components of cosmic radiation separately is also crucial to understand the individual effects of these components on organisms (Yi et al., 2013). Both approaches are valuable to understand the impact of cosmic radiation on biological systems and for understanding how various biological treatments can protect against the effects of cosmic radiation.

Although accelerator studies provide significant findings on how cosmic radiation affects biological organisms and how to increase protection, the experimental setups are different from the original scenario. Solar radiation can vary in both intensity and duration, lasting from several hours to days while galactic cosmic rays are continuous in the background with flux changes depending on the solar activity (Durante & Cucinotta, 2011). Replicating these conditions is not straightforward and requires more detailed information about the route and duration of the space mission. For example, a mission to Mars will take hundreds of days (Horneck et al., 2006) whereas a mission to the Moon will take several days (Märki, 2020). By using accelerators, acute exposure experiments can last up to a few hours, while chronic exposure experiments can continue for several weeks (Luitel et al., 2024). Ideally, experiments should closely resemble the space environment, where the flux and duration of cosmic ray exposure change over time (Durante & Cucinotta, 2011). This can be achieved by establishing new accelerator laboratories dedicated to space biology research or repurposing existing accelerators for specially using in this field (Durante & Kronenberg, 2005; Fogtman et al., 2023).

## 6. Conclusion

In this paper, we propose using biological response pathways, which have evolved to cope with Earthly stresses, for protecting living organisms from cosmic radiation in space. Studying these pathways could open the way for developing new drugs and specific diets for safe space travel. In addition to its direct harm, cosmic radiation can cause further harm by interacting with other objects (e.g., spacecraft and planetary surfaces) and produce secondary particles (Durante & Cucinotta, 2008). Therefore, it is essential to find the maximum protection against cosmic radiation. Interdisciplinary research in this field is a necessity for the safety of future space missions and the establishment of healthy extraterrestrial habitats (Furukawa et al., 2020). Thus, by combining the expertise of biologists, astrophysicists, particle/accelerator physicists, material scientists and space engineers, we can understand the damage caused by cosmic radiation better and develop protective biological strategies against it. We believe that this collaborative effort is critical for the success of space exploration in the future.

Finally, it is inevitable that mankind will eventually conquer near space in coming years and space radiation remains a significant obstacle. In our opinion, the launch of a new CERN-like scientific centre, the Dedicated International Accelerator Laboratory for Space Travel related radiation research (DIAL-ST), could make a significant contribution to advancing this critical area of research. This centre should include specially designed GeV-scale electron, proton and nucleus accelerators.

# Acknowledgements

ZS was supported by Leverhulme Trust Early Career Fellowship. We would like to thank Dr. Alexei Maklakov for his insightful comments and suggestions.